# Metastable Innershell Molecular State (MIMS)

Young K. Bae

Y.K. Bae Corporation, Tustin, California 92780

## Abstract

We propose that the existence of Metastable Innershell Molecular State (MIMS) was experimentally discovered by Bae et al. in hypervelocity (v > 100km/s) impact of nanoparticles. The decay of MIMS resulted in the observed intense soft x-rays in the range of 75 − 100 eV in agreement with Winterberg's recent prediction.

Corresponding Author: Young K. Bae, Ph.D.
Address: 218 W. Main St., Suite 102, Tustin, CA 92780, USA
e-mail: ykbae@ykbcorp.com
Phone: 714-838-2881
Fax: 714-665-8829

Corresponding Author Contact info: ykbae@ykbcorp.com, 714-838-2881

High energy density physics with an energy density greater than $10^{11}$ J/m$^3$ (equivalent pressure of 1 Mbar) is a rapidly growing field that spans a wide range of physics areas including plasma physics, laser and particle beam physics, nuclear physics, astrophysics, atomic and molecular physics, materials science and condensed matter physics, intense radiation-matter interaction physics, fluid dynamics, and magnetohydrodynamics. [1] Specifically, properties of matters under extreme conditions with energy densities in excess of $10^{13}$ J/m$^3$ (equivalent pressures greater than 100 Mbar) are important to understanding of extensive astrophysical phenomena and to realizing an efficient power generation by nuclear fusion.

The fundamental question is whether there are any new phenomena in the properties of matters under such extreme conditions. The specific internal energy, $\varepsilon$, and pressure, $p$, of a shock compression can be written as, $\varepsilon = \varepsilon_c + \varepsilon_i + \varepsilon_e$, and $p = p_c + p_i + p_e$, where $\varepsilon_c$ is the elastic energy due to repulsion in the electron potential curves, $\varepsilon_i$ is the ion thermal energy, $\varepsilon_e$ is the electron thermal energy, $p_c$ is the elastic pressure, $p_i$ is the pressure due to ion thermal motion, and $p_e$ is the pressure by electron thermal motion. The ratio, $\Gamma$, of the electron potential energy to the thermalization energy is defined as

$$\Gamma = \frac{\varepsilon_c}{\varepsilon_i + \varepsilon_e} = \frac{p_c}{p_i + p_e} \qquad (1)$$

We define here that when $\Gamma \leq 1$, the compression is defined "hot", and when $\Gamma \gg 1$, the compression is define "cold". At ultrahigh pressures in excess of 100 Mbar created by "hot" compression methods, the thermal energy becomes dominant, the role of the elastic components becomes small, and the material behaves practically as a perfect gas. [2] Therefore, in "hot" compression, the electronic binding and quantum effect are negligible factors.

Traditionally, in the laboratory scale setup, matters under the ultrahigh pressure have mostly been created and studied with "hot" compression methods, such as high-power short-pulse laser induced compression. [1] For example, during the laser induced compression, the laser beam energy heats electrons first, then ions are heated and compressed from thermalization of electron energy with ion energy. In such "hot" laser compression, no matter how short the laser pulse is, the ion compression time scale is thus limited by the electron-ion thermalization time scale of > 1 ps. [3] On the other hand, if an ultrahigh pressure compression has suppressed or minimal thermalization between ions and electrons, the initial compression energy can be mostly transferred to the elastic energy, and $\Gamma \gg 1$. Such a "cold" compression in the pressure range in excess of 100 Mbar has rarely been investigated, primarily because of the scarcity of experimental and theoretical studies. Matters under such "cold" ultrahigh pressure compression may have interesting quantum mechanical properties, which can not been accessed with "hot" compression.

We propose here that matter under "cold" compression with pressure in the order of 100 Mbar may generate Metastable Innershell Molecular State (MIMS) bound by innershell electrons with binding energy in excess of 100 eV in analogy to the generation of typical chemically bound (binding energy in the order of ~ 1 eV) molecular states at pressures in the order of 1 Mbar. We propose further that such "cold" compression can be readily generated in ultrahigh velocity (v>100km/s) impact of nanoparticles. For example, if the nanoparticles size is 1 – 10 nm in diameter, with the particle velocity of ~ 100 km/s, the collision time scale is 10 – 100 fs that is much shorter than that of ion-electron thermalization, which is in the order of a few ps. [3] When the nanoparticles impact on surface, initially most of kinetic energy is carried by ions.

Corresponding Author Contact info: ykbae@ykbcorp.com, 714-838-2881

Because the associated collision/compression time of the nanoparticles impact is much shorter than the ion-electron thermalization time, the initial kinetic energy of the nanoparticles will be mostly transferred to the elastic energy, $\varepsilon_c$. During the initial stage of impact in the time scale of 10 – 100 fs, the nature of the nanoparticle compression should thus be very similar to that of "cold" compression with $\Gamma \gg 1$. Therefore, the proposed nanoparticle impact provides a new pathway to investigating high energy density matter under the "cold" ultrahigh pressure compression.

The potential existence of MIMS in highly compressed plasma with relatively low temperature was numerically investigated and predicted by Younger et al. [4] They performed a set of prototypical quantum calculations of the electronic structure of a nine-atom helium plasma over wide ranges of temperature and density. The results of the calculations revealed the presence of very tightly bound quasimolecular states, MIMS, in high-density plasmas, even at temperatures high enough to ionize fully the component atoms. They found that such quasimolecular quantum states, MIMS, gradually disappear as temperature increases. The nature and optical or quantum properties of such excited states, however, were not investigated.

More recently, Winterberg [5,6] predicted that if matter is "suddenly" put under a high pressure in excess of 100 Mbar, it can undergo a transformation into MIMS with keV potential wells for the electrons. Further, Winterberg [5,6] predicted that the decay of MIMS can emit extremely intense x-rays. Based on the calculations done by Muller et al, [7] Winterberg further predicted the x-ray energy, $\delta E$, can be approximately estimated by [6],

$$\log_{10}(\delta E) \cong 1.3 x 10^{-13} Z - 1.4 \qquad (2)$$

where Z is the sum of the nuclear charge for both components of the molecule formed under the high pressure in the order of 100 Mbar.

If MIMS exists, it would decay into the lower orbit via radiative and non-radiative processes. The radiative lifetime, $\tau_{rad}$, is given by [8]

$$\tau_{rad} \approx \frac{4.5 \times 10^{-8} \lambda^2}{n} , \qquad (3)$$

where $\lambda$ is the wavelength of the radiation in $\mu m$ and n is the index of refraction. For example, if the MIMS emits x-ray photons with an energy of 1 keV, $\lambda = 1.2 x 10^{-3}$ $\mu m$, $\tau_{rad} = 6.5 x 10^{-14}$ s. Winterberg further predicted that the photon flux, $\phi$, of the radiative decay of the excited states, is given by [5,6], $\phi \approx \frac{cp}{6}$, where c is the speed of light and p is the pressure. For example, for p=100 Mbar, $\phi=5 x 10^{15}$ W/cm$^2$.

We propose that MIMS and the associated intense x-ray radiation [5,6] can be more readily produced in "cold" compression in nanoparticles hypervelocity (v>100km/s) impact. As it was mentioned above, the optical decay MIMS formed by heavy element atoms can be fast enough to compete with other non-radiative decay channels. However, the optical decay MIMS formed by light element atoms may not be fast enough to compete with other non-radiative decay channels. For example, if the associated radiation energy of the light element atom MIMS decay is 0.1 keV, $\lambda = 1.2 x 10^{-2}$ $\mu m$, $\tau_{rad} = 6.5 x 10^{-12}$ s, which is much larger than the ion-electron thermalization time scale and other non-radiative decay lifetimes. In this case, we show here that the hypervelocity impact of nanoparticles can increase the rate of optical decay of MIMS by orders of magnitude via Dicke superradiance mechanism, [9] in which the size of compressed

Corresponding Author Contact info: ykbae@ykbcorp.com, 714-838-2881

particle can be smaller than that the wavelength of the associated radiation. R. H. Dicke [9] predicted that if N quantum oscillators prepared in inverted but incoherent fashion are all contained in a volume small compared to the emission wavelength cubed, the quantum oscillators would all be coupled together through their overlapping radiation fields.[9] In Dicke superradiance, the radiation will be emitted in a burst with duration of $\tau_{rad}$/N. [9,10] For example, in a 0.1 keV photon radiation decay of nanoparticles (N~100), with the isolated individual excited molecular states with a radiative decay lifetime, $\tau_{rad} = 6.5 \times 10^{-12}$ s, the Dicke superradiance drastically decrease the radiative decay lifetime to ~ $\tau_{rad}$/N~ $6.5 \times 10^{-14}$ s, which is comparable with the collision time and any other thermalization times. Therefore, during the nanoparticle collision the radiative process can be a dominant decay process of the proposed energetic excited molecular states.

If the number of atoms in nanoparticles is large enough, the impact generates hydrodynamic shock behavior that can be described by one dimensional shock theory. In this case, [2] the pressure $P_s$ and the density $\rho_s$ of the shocked target material in a one-dimensional strong shock produced by the particle impact at a velocity v are:

$$P_s = \frac{4}{3}\rho_P v^2, \tag{4}$$

$$\rho_S = \frac{(\gamma+1)}{(\gamma-1)}\rho_T. \tag{5}$$

Hypervelocity "cold" compression can have $\gamma$ much different from 5/3, which is for a monoatomic non-ionized gas. The exact value of $\gamma$ is crucial in determining the $\rho_S$, and depending on material characteristics, $\rho_S$ can be greater than $8\rho_T$. [2] In this case, the interatomic distance during the collision would be shorter than ½ of the normal distance as in the Winterberg's case. [5,6]

We show that the intense x-rays produced in Dicke superradiance of MIMS triggered the anomalous particle detector response to the hypervelocity impact of nanoparticles, which was discovered by Bae et al. [11,12]. More specifically, in their experiments, Bae et al. [11,12] discovered that a passivated solid state detector with a 500 A boron doped window was observed to be sensitive to the impact of the nanoparticles although the nano-articles have isolated atomic ranges much smaller than the thickness of the detector window. [11] With the use of this discovery, Bae et al. [11] measured stopping powers of large multiply charged ions by determining energy losses in thin aluminum films coated on the solid state detectors. With initial projectile velocities in the range of 100 – 500 km/sec, the energy losses of these clusters and molecules were observed to follow more hydrodynamic shock behavior than the individual atomic stopping power behavior. The transient pressures estimated during the impact using a one-dimensional shock model were estimated to be in excess of 100 Mbar. [11]

To understand the discovery further, Bae et al. performed a subsequent work in which the response of the passivated solid state detector to the impact water cluster ions with $(H_2O)_N H^+$ with N from 1 to 1500 was systematically investigated at an acceleration energy of 500 keV. [12] The data with 1<N<10 were in agreement with the theoretical curves calculated with the energy loss through the 500 A window based on individual atom's stopping power. This agreement ruled out the possibility that the window may be partially sensitive to the particle impact. A surprising aspect of the experimental results was that the passivated solid-state detector with the window thickness of 500 A responded to the impacts of large molecules and clusters that have isolated atomic ranges up to 30 times smaller than the window thickness.

Corresponding Author Contact info: ykbae@ykbcorp.com, 714-838-2881

The anomalous detector response may result from the energetic ions or electron shower produced in shocks created by the impact of nanoparticles. The detector response to the nanoparticle impact was not observed through a 2,300-A thick Al window directly deposited on a bare detector, thus the upper bound energy of these electrons is estimated to be ~ 3.5 keV. [12] Bae et al. investigated whether the pulse height of the 500 A window detector changed upon tilting the detector by 45° thereby increasing the window thickness by a factor of 1.41, or from 500 A to 700 A. The energies through 500 A and 700 A thick Si windows of 3.5 keV electrons would be 3.1 keV and 2.8 keV, respectively. Such a change would be readily detected in their experiment, however, the absence of change ruled out the possibility of intense electron shower. [13] If the energetic ions are protons, their upper energy is 20 keV. The energy through 500 A and 700 A thick Si windows of 20 keV protons would be 10 keV and 8 keV, respectively. [14] Such a change would be readily detected in the experiment by Bae et al. [12] however, the absence of the change ruled out the possibility of energetic proton ions as well. Because ions heavier than protons would have more pronounced energy loss through the window, the possibility of energetic ions heavier than protons is ruled out as well. Therefore, the only possible explanation for the detector sensitivity seems generation of intense x-rays during the impact. Because there was no changes observed by tilting the detector, the Si window with thickness from 500 – 700 A should be relatively transparent to the observed x-rays.

Here, we have analyzed in depth of the water cluster data with N>50 by Bae et al. (presented in the Fig. 2 of Bae et al. [12]) in terms of pressure and particle energy per water molecules in Fig. 1. The pulse height, y-axis, represents the ratio of the cluster energy detected through the 500 A window to the single water molecular ion (N=1) at 500 keV in percent. The x-axis also represents the approximate shock pressure, $P_s$, generated by water cluster impact from Eq. 4. The detector pulse height with 500 keV N=1 ions should result mostly from the electronic stopping power. Because the radiation energy should only be related with electronic energy in the detector, the pulse height comparison between the pulse heights from single water cluster impact and the radiation generated by large water clusters is a reasonable measure for estimating the radiation energy.

The experimental data in Fig. 1 show a possible existence of an activation energy, thus were curve fitted with the Arrhenius Equation, $H = H_0 \exp(-\dfrac{E_a}{E})$, where H is the relative pulse height, $H_0$ is a constant, $E_a$ is the reaction threshold kinetic energy per water molecule, and E is the impact kinetic energy per water molecule. The pulse height slightly deviates at high impact energies beyond 10 keV with the cluster size smaller than 50, which are expected to be less efficient in generating shock waves. The curve fit is excellent, and resulted in $H_0$=19, and $E_a$ = 0.15 keV. At this threshold kinetic energy the equivalent threshold velocity of impact is 40 km/sec, and the threshold impact pressure is 21.3 Mbar.

Another surprising aspect of the data was that the detector sensitivity of large clusters (N>50) was as much as ~20 % of that of N=1 ions at the same total kinetic energy. If the radiation process generated from the optical decay of the energetic molecular state during impact is symmetric along the plane parallel to the target surface, the detector would collect only ½ of the total radiation, and the pulse height represents 1/2 of the total radiation energy. This is illustrated in Fig. 1; The right side y-axis represents for the fraction of radiation during the impact in percent. The maximum radiation fraction is as high as 38 %.

Fig. 2 of Bae et al. [11] presented the data with water cluster impact obtained with a ruggedized detector with a 2300-A thick Al window. The data with the detector with 2300-A thick Al window show that the detector was insensitive to the ions with n>7 in agreement with the calculated ranges. Therefore, it seems that the radiation is fully absorbed through 2300 A Al window. To understand this effect further, we calculated and plotted the transmission [14] of x-ray photons through 2300 A aluminum film on the detector in Fig. 2. The 2300 A aluminum coating would fully absorb soft x-rays with the energy of 75-170 eV. Therefore, if the radiation energy is mostly from the x-ray, their energy is estimated to be 75 -170 eV. The observed transparency through 500-A thick Si windows in the experiment with tilted detectors can be explained if the x-ray energy is below ~100 eV below the threshold of 2p subshell photoionization process of Si. [15] By combining the results obtained with the 2,300-A thick Al film and the 500-A thick Si window, the x-ray range is estimated to be 75 – 100 eV.

The predicted radiation energies with Eq. 2 derived by Winterberg [6] of different combination of atoms are listed in Table 1. Because the main kinetic energy of water clusters is carried by oxygen atoms, the x-ray induced water clusters should come mainly from O - O, O – Si and Si – Si for water cluster impact on Si, and O – O, O – Al and Al – Al for water cluster impact on Al. The agreement with the experimental data on the energy range of 75 – 100 eV by Bae et al, [12] and the theoretically predicted radiation energy by Winterberg's equation [6] is excellent. The detailed quantum mechanical energy level calculation should be performed to increase the accuracy of the energy level of the excited states radiation energy.

Fig. 3 shows p - d (pressure-lattice distance) diagram, proposed by Winterberg, of water cluster impacting Al or Si surface. This diagram illustrates the formation of the excited molecular states during the compression along the adiabatic curve A at the critical distance of $d_{cr}$, where the pressure reaches the critical pressure, $p_{cr}$. In passing over the critical pressure the excited molecular state decay into the lower state by emitting x-ray photons. After the x-ray emission, the molecule disintegrates along the lower adiabat B. From the previous analysis of the data by Bae et al., [11,12] it is estimated that $p_{cr} \sim 21$ Mbar, and $p_r$ in the range of $7 – 10$ Mbar.

The radiation from Winterberg's process [5,6] for N~200 water cluster impact with R~8x10$^{-8}$ cm, is considered further here. Assuming the radiation energy, $\tau_{rad}$, for the excited molecular states with the radiation energy of 90 eV ($\lambda \sim 0.012$ $\mu$m) is 7.6 x 10$^{-12}$ s, which is much larger than the collision time scale. However, with N~200, the corresponding Dicke superradiance time width is 7.6 x 10$^{-12}$/200~ 3.8x10$^{-14}$ s, which is in the order of the collision time. As N increase, the size of the cluster becomes comparable with the wavelength of the radiation, then Dicke superradiance effect will be reduced. This will happen near the cluster size of 12 nm for $\lambda \sim 12$ nm radiation, which is equivalent to N~100,000. Therefore, the investigated cluster size range is well below the maximum allowed cluster size for Dicke superradiance. [9,10]

If during collision the diameter of cluster is decreased by a factor 2, the surface area, A, during collision is 2x10$^{-14}$ cm$^2$. If we assume the collision time, $t_c$, is in the order of 10$^{-14}$ s, and the radiation time is similar, the total photon energy, $E_{ph}$, from the N~200 water cluster is given by $E_{ph} \sim \phi A t_c$. Assuming the pressure of compression is ~ 100 Mbar, $\phi \sim 5x10^{16}$ W/cm$^2$, $E_{ph} \sim 1x10^{-11}$ J ~ 6x10$^7$ eV. The observed radiation energy is in the order of ~ 2 x 10$^5$ eV (~38 % of 500 keV), thus it seems that Winterberg's [5,6] formula overestimates the total radiation energy by more than two orders of magnitude. Currently, we do not know what the cause of this discrepancy is.

The other interesting aspect of the data by Bae et al. [11] was that the detector sensitivity of the large biomolecules of the through thin Al films decreased logarithmically as a function of

Corresponding Author Contact info: ykbae@ykbcorp.com, 714-838-2881

the film thickness. If these biomolecules generated the soft x-rays with photon energies of 70 - 100 eV through these thin Al films, because the Si detector window is essentially transparent to the soft x-rays, the sensitivity decreases would be too small to be observed in their experiment. However, the experiment with thin films can have another process in competition with the excitation of the energetic molecular state; Energetic particles passing through thin films can excite plasmons. [16]

The excited plasmons decay mainly via phonon coupling and optical coupling. [17] Phonon coupled decay of plasmons would not generate any signals in the silicon detector, however the optical coupling may. In fact due to Dicke superradiance effect, the optical decay of plasmons in nanoparticle impact regions may be amplified by a factor of N (the number of involved atoms in the impact). [17] For the photon energy of Al film is ~15 eV (volume plasmon), the radiation decay lifetime of the plasmons is $\tau_{rad} \sim 2.9 \times 10^{-10}$ s, which is much larger than other decay processes. For albumin impact, N is ~ 5,000. With the effect of Dicke superradiance, the optical decay lifetime of the excite phonons increases to $\tau_{rad}/N \sim 5.8 \times 10^{-14}$ s, which is in the order of the collision time. Therefore, the passage of albumin molecules through Al films would generate intense burst of 15 eV photons. Because of the Si optical absorption characteristics, [15] the 15 eV will be completely attenuated through the 500-A thick Si window, thus the photons generated in the Al films would not generate any detector signals in agreement with the observed data by Bae et al. [11]

To understand the detector sensitivity more in depth, similar fitting procedures to the water cluster impact data were applied to albumin[47+] and cytochrome-c[18+] data by Bae et al. (presented in the Fig. 5 of Bae et al. [12]). Because the estimated densities of albumin[47+] and cytochrome-c[18+] are ~1/10 of the water cluster, their data can be used to probe into the effect of the nanoparticle density variation on the shock development. [12] The results of fitting of albumin[47+] and cytochrome-c[18+] data to Arrhenius equation are excellent. Various relevant parameters and fitting results for water clusters and the above two molecules are summarized in Table 2.

Although the density of molecules is changed by a factor of 12 by switching water molecules to albumin[47+], the estimated critical pressure, $P_{cr}$, increased by a factor of 1.5. For cytochrome-c[18+], $P_{cr}$ increased by a factor of 2.8. Because of their low density, both albumin[47+] and cytochrome-c[18+] impacts are expected to have some characteristics of atomic impact. In this case, during impact the projectile atoms would penetrate deeper into the target, and the number of atoms in the impact region generating the compression could be considerably larger than the number of atoms in the projectile molecules due to the involvement of target atoms in the collision process. In such cases, $E_{cr}$ can be increased and the critical pressure estimated from $E_{cr}$ may overestimate $P_{cr}$. Therefore, $P_{cr}$ data derived from albumin[47+] and cytochrome-c[18+] data represent the upper bounds of $P_{cr}$, and the actual $P_{cr}$ reflecting these effects for all these three projectiles may be similar. If this is the case, $P_{cr}$ is a slowly varying function of the projectile density.

Finally, we propose that the proposed intense x-ray production with nano-particle impact can be used to generate intense hard x-rays. The rate of radiative decay time is proportional to $\lambda^2$ as in Eq. 3. For example, according to Winterberg [5,6], MIMS formed by [92]U -[92]U pairs, for example, would emit ~ 10 keV x-rays with $\lambda = 1.2 \times 10^{-4}$ μm, $\tau_{rad} = 6.5 \times 10^{-16}$ s. Therefore, efficient x-ray generation from matters composed of heavy elements may not require the usage of nanoparticles and Dicke superradiance mechanism. Based on the experimental observation by Bae et al. [11,12] and present analyses, the kinetic energy per atom required for triggering the x-ray production mechanism is in the order of the x-ray energy. For the $^{92}U$ -$^{92}U$ pairs, the required velocity of $^{92}U$ nanoparticles to achieve such threshold energy is ~ 100 km/sec, of which corresponding threshold pressure is ~ 2 Gbar.

Although Dicke superradiance mechanisms may not be important in producing intense x-ray radiation for impact of nanoparticles consisting of heavy elements, such as $^{92}U$, still there is another factor limiting the size of the particles, which is the ion-electron thermalization time scale of 1 ps. [3] The maximum nanoparticle diameter for producing the intense x-rays at impact velocities of ~100 km/s is ~ 100 nm, which is much larger than the size limitation by Dicke superradiance of 0.12 nm.

In sum, we propose that Metastable Innershell Molecular State (MIMS) can be readily created in "cold" compression with pressures in excess of 100 Mbar and that such "cold compression can be generated in the hypervelocity (v>100 km/s) of nanoparticles, in which the collision/compression time scale (10-100 fs) is shorter than the ion-electron thermalization time scale (>1 ps). Further, we propose here that the limited size of nanoparticles can increase the emission rate of MIMS x-rays owing to the Dicke superradiance mechanism. Our theory combined with the Winterberg's recent prediction explains that the anomalous detector signals discovered by Bae et al. in hypervelocity (v > 100 km/s) impact of nanoparticles, such as clusters and biomolecules, resulted from the existence and optical decay of MIMS. The analysis of the experimental data resulted in the energy of intense soft x-rays in the range of 75 – 100 eV in agreement with Winterberg's prediction, and the conversion efficiency of 38 % from the initial kinetic energy of nanoparticles to the x-ray radiation energy.

**Acknowledgements**
    We thank Prof. F. Winterberg for stimulating discussions.

Corresponding Author Contact info: ykbae@ykbcorp.com, 714-838-2881

Corresponding Author Contact info: ykbae@ykbcorp.com, 714-838-2881

**Table 1. Predicted soft x-ray energy, δE, by Winterberg [5,6] vs experimental energy range data by Bae et al.**

|  | O - O | O - Si | Si - Si | O − Al | Al - Al |
|---|---|---|---|---|---|
| δE (eV) | 64 | 77 | 92 | 75 | 87 |
| Experimental (eV) | 75 - 100 | | | | |

**Table 2. Various properties and fitting results to Arrhenius equation of nanoparticles investigated by Bae et al. {11,12} The fittings are excellent and $P_{cr}$ is estimated from Eq. 4.**

|  | Mass (amu) | $E_{cr}$ (keV/amu) | Size (diameter in A) | Density (g/cc) | $P_{cr}$ (Mbar) |
|---|---|---|---|---|---|
| $(H_2O)_NH^+$ | 900-27,000 | 0.0083 | 12 - 44 | 1.0 | 21 |
| albumin[47+] | 66,400 | 0.14 | 136 | 0.085 | 30 |
| cytochrome-c[18+] | 12,400 | 0.23 | 72 | 0.1 | 58 |

Corresponding Author Contact info: ykbae@ykbcorp.com, 714-838-2881

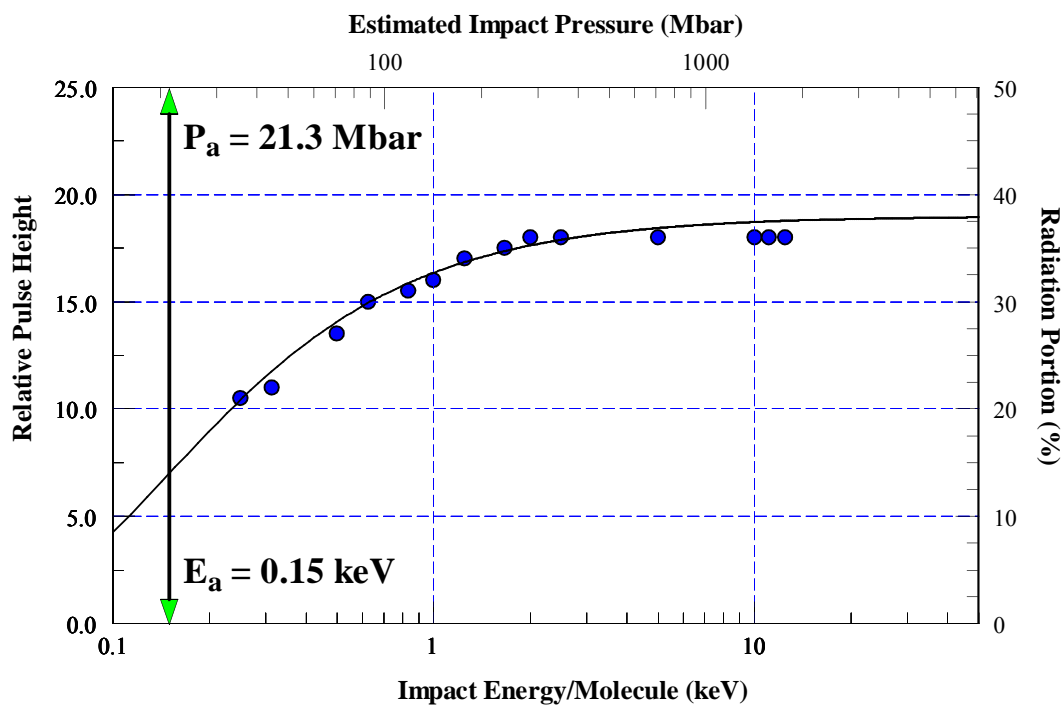

**Figure 1.** The filled circles represent the Relative Pulse Height of the detector response with a silicon detector with a 500 A thick window as a function of cluster Impact Energy per water molecule obtained by Bae et al. [12] The solid curve represents the best fit of the data with the Arrhenius equation. The fitting is excellent, and resulted in the threshold energy and pressure of Ea = 0.15 keV and 21.3 Mbar respectively.

Corresponding Author Contact info: ykbae@ykbcorp.com, 714-838-2881

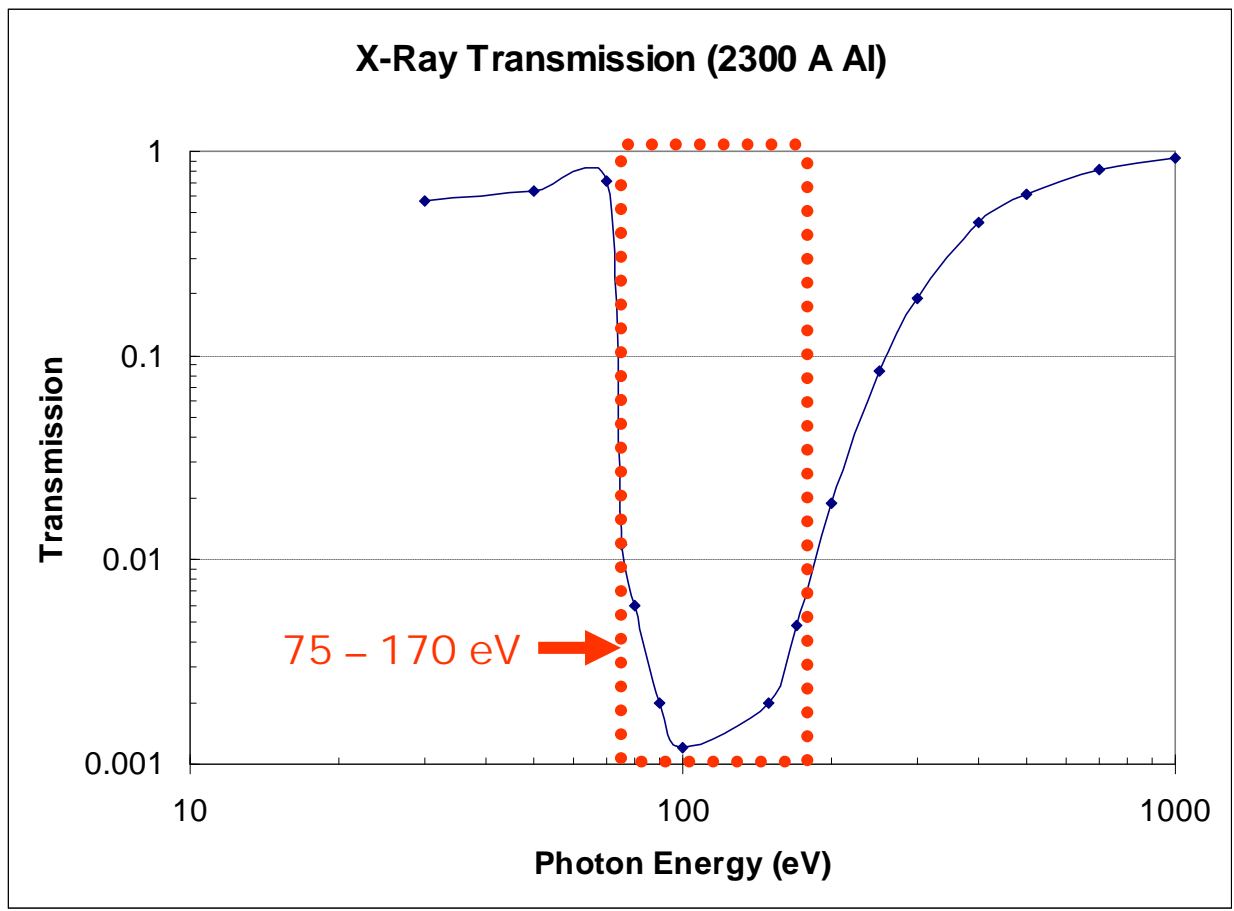

**Figure 2.** **Optical transmission through a 2300 A Al film. [13] The dotted rectangle (75 – 170 eV) represents the absorption range of the Al film by more than a factor of 100, which defined the threshold of detectable signals of the Si detector by Bae et al. [11,12]**

Corresponding Author Contact info: ykbae@ykbcorp.com, 714-838-2881

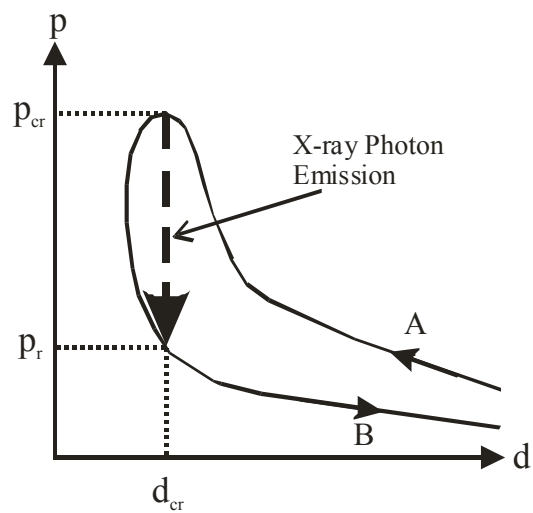

**Figure 3. p - d (pressure-lattice distance) diagram, proposed by Winterberg, [6] of water cluster impacting a Si surface. From the analysis of Bae et al., [11,12] it is estimated that $p_{cr} \sim 21$ Mbar, and $p_r$ in the range of 7 – 10 Mbar with soft x-ray photon emission with energy of 75-100 eV at a conversion efficiency from the kinetic energy to photon energy of ~38 %.**

Corresponding Author Contact info: ykbae@ykbcorp.com, 714-838-2881